\documentclass[psfig]{aa}
\input psfig.tex

\begin{document}                                                                

\twocolumn


   \title{The REFLEX Galaxy Cluster Survey\,VII: $\Omega_m$ and
         $\sigma_8$ from cluster abundance and large-scale clustering}

   \titlerunning{The REFLEX Galaxy Cluster Survey}

   \author{Peter\,\,Schuecker\,$^{(1)}$, Hans\,\,B\"ohringer\,$^{(1)}$,
           Chris\,A.\,\,Collins\,$^{(2)}$ and Luigi\,\,Guzzo\,$^{(3)}$}

   \authorrunning{Schuecker et al.}

   \offprints{Peter Schuecker\\ peters@mpe.mpg.de}

   \institute{
    $^{(1)}$ Max-Planck-Institut f\"ur extraterrestrische Physik,
             Giessenbachstra{\ss}e 1, 85740 Garching, Germany\\ 
    $^{(2)}$ Astrophysics Research Institute, Liverpool John Moores
University, Twelve Quays House, Egerton Wharf, Birkenhead CH41 1LD,
Great Britain\\
    $^{(3)}$ Osservatorio Astronomico di Brera, via Bianchi, 22055
Merate (LC), Italy\\
}
                                                                                
   \date{Received  ; accepted }                         
   
   \markboth{The REFLEX Galaxy Cluster Survey}{}

\abstract{For the first time the large-scale clustering and the mean
abundance of galaxy clusters are analysed simultaneously to get
precise constraints on the normalized cosmic matter density $\Omega_m$
and the linear theory RMS fluctuations in mass $\sigma_8$. A
self-consistent likelihood analysis is described which combines, in a
natural and optimal manner, a battery of sensitive cosmological tests
where observational data are represented by the (Karhunen-Lo\'{e}ve)
eigenvectors of the sample correlation matrix. This method breaks the
degeneracy between $\Omega_m$ and $\sigma_8$. The cosmological tests
are performed with the ROSAT ESO Flux-Limited X-ray (REFLEX) cluster
sample. The computations assume cosmologically flat geometries and a
non-evolving cluster population mainly over the redshift range
$0<z<0.3$. The REFLEX sample gives the cosmological constraints and
their $1\sigma$ random errors of
$\Omega_m\,=\,0.341\,^{+0.031}_{-0.029}$ and
$\sigma_8\,=\,0.711\,^{+0.039}_{-0.031}$. Possible systematic errors
are evaluated by estimating the effects of uncertainties in the
value of the Hubble constant, the baryon density, the spectral slope
of the initial scalar fluctuations, the mass/X-ray luminosity relation
and its intrinsic scatter, the biasing scheme, and the cluster mass
density profile. All these contributions sum up to total systematic
errors of $\sigma_{\Omega_m}=^{+0.087}_{-0.071}$ and
$\sigma_{\sigma_8}=^{+0.120}_{-0.162}$.  }

\maketitle

\section{Introduction}\label{INTRO}

Several observational arguments suggest that we live in a
geometrically flat Universe in a phase of accelerated cosmic expansion
(Riess et al. 1998, Perlmutter et al. 1999, Stompor et al. 2001,
Netterfield et al. 2002, Pryke et al. 2002, Sievers et
al. 2002). Strong indications for a low cosmic matter density come
from, e.g., the abundance of galaxy clusters (e.g., Carlberg et
al. 1996, Viana \& Liddle 1996, Bahcall \& Fan 1998, Eke et al. 1998,
Henry 2000, Borgani et al. 2001, Reiprich \& B\"ohringer 2002,
hereafter RB02), and from the large-scale distribution of clusters
(e.g. Collins et al. 2000, Schuecker et al. 2001, hereafter Paper\,I)
and galaxies (e.g., Percival et al. 2001, Szalay et al. 2001). A
recent overview including also new results on gravitational dynamics,
weak lensing and the baryon mass fraction in clusters of galaxies can
be found in Peebles \& Ratra (2002).

With the present investigation we are improving the constraints on the
matter density $\Omega_m$ and the linear theory RMS matter
fluctuations $\sigma_8$ in comoving spheres with $8\,h^{-1}\,{\rm
Mpc}$ radius. We base our approach on observational estimates of
spatial clustering and abundance of galaxy clusters. Simultaneous
constraints on $\Omega_m$ and $\sigma_8$ are derived under the
assumption of a flat geometry of the Universe. Cluster abundances
although almost insensitive to the cosmological constant
$\Omega_\Lambda$ are known to be quite sensitive to $\Omega_m$ and
actually one of the best ways of measuring $\sigma_8$ (White et
al. 1993, Eke et al. 1996, Kitayama \& Suto 1997, Borgani et al. 1997,
Mathiesen \& Evrard 1998). Our method uses the statistical information
of the present epoch large-scale structure to characterize the
cosmological model. In comparison to other methods of the assessment
of the large-scale structure e.g. through the galaxy distribution, the
analysis of the galaxy cluster distribution has the advantage that it
relies on astrophysics mostly governed by gravitation, so that cluster
masses and biasing factors can be obtained from first principles,
especially when the clusters are selected in X-rays (e.g., Borgani \&
Guzzo 2001).

We use the ROSAT ESO Flux-Limited X-ray (REFLEX) galaxy cluster
sample. In the REFLEX survey special care was taken to get a
well-controlled sample selection and a high completeness (B\"ohringer
et al. 2001). The sample consists of the 452 X-ray brightest southern
clusters with redshifts mainly below $z=0.3$, selected in X-rays from
the ROSAT All-Sky Survey (RASS, Voges et al. 1999) and is confirmed by
extensive optical follow-up observations within a large ESO Key
Programme (B\"ohringer et al. 1998, Guzzo et al. 1999). The sample has
been used to measure with unrivalled accuracy the cluster X-ray
luminosity function (B\"ohringer et al. 2002), the spatial
cluster-cluster correlation function (Collins et al. 2000), and its
power spectrum (Paper\,I).

In order to get accurate constraints on $\Omega_m$ and $\sigma_8$ we
use several independent cosmological tests based on the dependence of
the cosmic matter power spectrum and the volume on the values of the
cosmological parameters. On the one side we measure the spatial
fluctuations of galaxy clusters. This test is known to be especially
sensitive to the shape of the matter power spectrum on large
scales. On the other side we measure the average abundance of
clusters. This test is known to be especially sensitive to the
normalization of the matter power spectrum on small scales. Finally
and independent from the matter power spectrum, we count clusters to
measure their abundance as a function of redshift and thus to measure
the volume which again strongly depends on cosmology (redshift-volume
test, e.g., Robertson \& Noonan 1968).

A natural way to combine the different tests is offered by the
Karhunen-Lo\'{e}ve (KL) eigenvector analysis. The method was first
applied by Bond (1995) to analyse cosmic microwave background (CMB)
temperature maps and translated by Vogeley \& Szalay (1996) to the
case of the spatial analysis of galaxies. The KL eigenvectors $\Psi$
are constructed in a way to obey orthogonality properties which allow
unbiased studies of fluctuations up to the largest scales covered by a
survey. The method avoids artifical correlations introduced by the
survey window between different fluctuation modes, affecting under
realistic conditions all power spectral analyses based on plane-wave
expansions on large scales. Applications of the KL method to galaxy
surveys can be found in Matsubara, Szalay \& Landy (2000) and Szalay
et al. (2001). Their KL tests are, however, still restricted to the
analysis of the clustering properties of galaxies.

To determine a precise eigenvector base $\Psi$ which does not
bias the final results, a more refined fiducial cosmological model has
to be assumed which is already close to our best model fits (the
construction of the fiducial model is described in Sect.\,\ref{PRAC}):
a pressure-less spatially flat Friedmann-Lema\^{\i}\-tre model, the
cosmic matter density $\Omega_m=0.3$, the cosmological constant
$\Omega_\Lambda=0.7$, the Hubble constant in the form $h=H_0/100\,{\rm
km}\,{\rm s}^{-1}\,{\rm Mpc}^{-1}=0.7$, the linear RMS normalization
$\sigma_8=0.75$, the spectral index of initial scalar fluctuations
$n_{\rm S}=1.0$, the baryon density $\Omega_bh^2=0.020$, and the CMB
temperature $T_{\rm CMB}=2.728$\,K.

In our first KL study of the REFLEX sample we followed the approach of
A.\,Szalay and collaborators, applying the KL method to the analysis
of three-dimensional spatial fluctuations (Schuecker et al. 2002,
hereafter Paper\,II).  The resulting constraints on the matter
density, $0.03<\Omega_mh^2<0.19$ (95\% confidence interval), were
found to be quite robust against major changes in the model
assumptions.

The present investigation completes our previous KL study by utilizing
now both the spatial fluctuations of galaxy clusters and their mean
abundance to get constraints on $\Omega_m$ and $\sigma_8$.  Some
general features of the KL method relevant for the present work are
outlined in Sect.\,\ref{METHOD}. In Sect.\,\ref{MODEL} we briefly
describe a model for the matter and cluster power spectrum already
developed in Paper\,II. We further introduce a new theoretical model
for the average cluster abundance which replaces the more
phenomenological prescription used in Paper\,II. The REFLEX sample and
the values of important model parameters are discussed in
Sect.\,\ref{PRAC}. Sect.\,\ref{RESULTS} summarizes the cosmological
constraints obtained with the REFLEX sample. In
Sect.\,\ref{CONCLUSION} we discuss the results and draw our basic
conclusions. In the Appendix we derive equations used to relate
cluster masses defined for different density contrast and background
levels.

\section{General method}\label{METHOD}

Clusters of galaxies are counted in cells. The results are summarized
in a vector $\vec{D}$ with elements containing the counts obtained
within the individual cells. Similarily, a second vector $\vec{N}$ is
introduced which contains the average (expected) model counts. The
latter are the diagonal elements of a noise matrix $N$. After the
equalization of the different noise in the cells by $N^{-1/2}$
(whitening) and the KL transformation into the $\Psi$ vectorbase, the
values of the cosmological parameters are estimated by minimizing the
differences $\Delta\vec{B}\,=\,\vec{B}\,-<\vec{B}>$ between the KL
coefficients $\vec{B}\,=\,\Psi^{\rm T}\,N^{-1/2}\,\vec{D}$ determined
by the observed sample, and the expected KL coefficients
$<\vec{B}>\,=\,\Psi^{\rm T}\,N^{+1/2}$ determined by a cosmological
model (see Paper\,II), where T denotes the transpose of a matrix.

More specifically, the columns of the unitary matrix $\Psi$ are the KL
eigenvectors of the so-called whitened correlation matrix
$R'\,=\,N^{-1/2}\,R\,N^{-1/2}$, with the elements of $R$ given by
$R_{ij}\,=\,N_i\,N_j\xi_{ij}\,+\,\delta_{ij}N_i$ and
$\xi_{ij}\,=\,\frac{1}{V_i\,V_j}\,
\int_{V_i}d^3\vec{r}_i\,\int_{V_j}d^3
\vec{r}_j\,\xi(|\vec{r}_i-\vec{r}_j|)$. Here, $\xi$ is the cluster
correlation function, and $V_i$ the volume of the $i$-th count cell
centered at the comoving coordinate $\vec{r}_i$.

The elements of $\Delta\vec{B}$ show random cell-to-cell fluctuations
caused by both the Poisson count noise and the large-scale
distribution of the clusters. For the correct weighting of
$\Delta\vec{B}$ these fluctuations are taken into account. Note that
the analysis of the fluctuations alone already gives useful
cosmological constraints as described in our first KL analysis
(Paper\,II).

A simple way to perform the minimization which also utilizes the
cosmological information contained in the random fluctuations between
the cells is suggested by the observed frequency distribution of the
KL coefficients. For the REFLEX cluster survey we found (Paper\,II)
that for large cell sizes, $\vec{B}$ follows a Gaussian distribution
with high statistical significance (new results are given in
Fig.\,\ref{FIG_DB} of the present paper). A multivariate sample
likelihood function of an assumed cosmological model should thus have
the form
\begin{equation}\label{M2}
{\cal L}(\vec{B}|\vec{x})\,=\,
\frac{\exp\left[-\frac{1}{2}\,{\rm
trace}\left(C^{-1}Z\right)\right]}{\sqrt{(2\pi)^{M}|{\rm
det}C|}}\,,
\end{equation}
with the sample covariance matrix $Z=\Delta\vec{B}(\Delta\vec{B})^{\rm
T}$, the model covariance matrix of the KL coefficients
$C\,=\,\Psi^{\rm T}\,R'_{\rm model}\,\Psi$, and the parameter vector
$\vec{x}$ specifying the cosmological model. We assume as usual that
the maximum of (\ref{M2}) determines the values of the model
parameters which yield the highest probability of obtaining cluster
abundances and fluctuations as large as observed. Note that in
Paper\,II, $Z$ was fixed by a phenomenological background model
(independent from any cosmological parameter) and only $C$ was
regarded as a free model-dependent variable. In the present
investigation both $Z$ and $C$ are computed with observed and modelled
data in a consistent manner (see Sec.\,\ref{MODEL}).

\begin{figure*}
\vspace{-0.0cm}
\centerline{\hspace{-10.5cm}
\psfig{figure=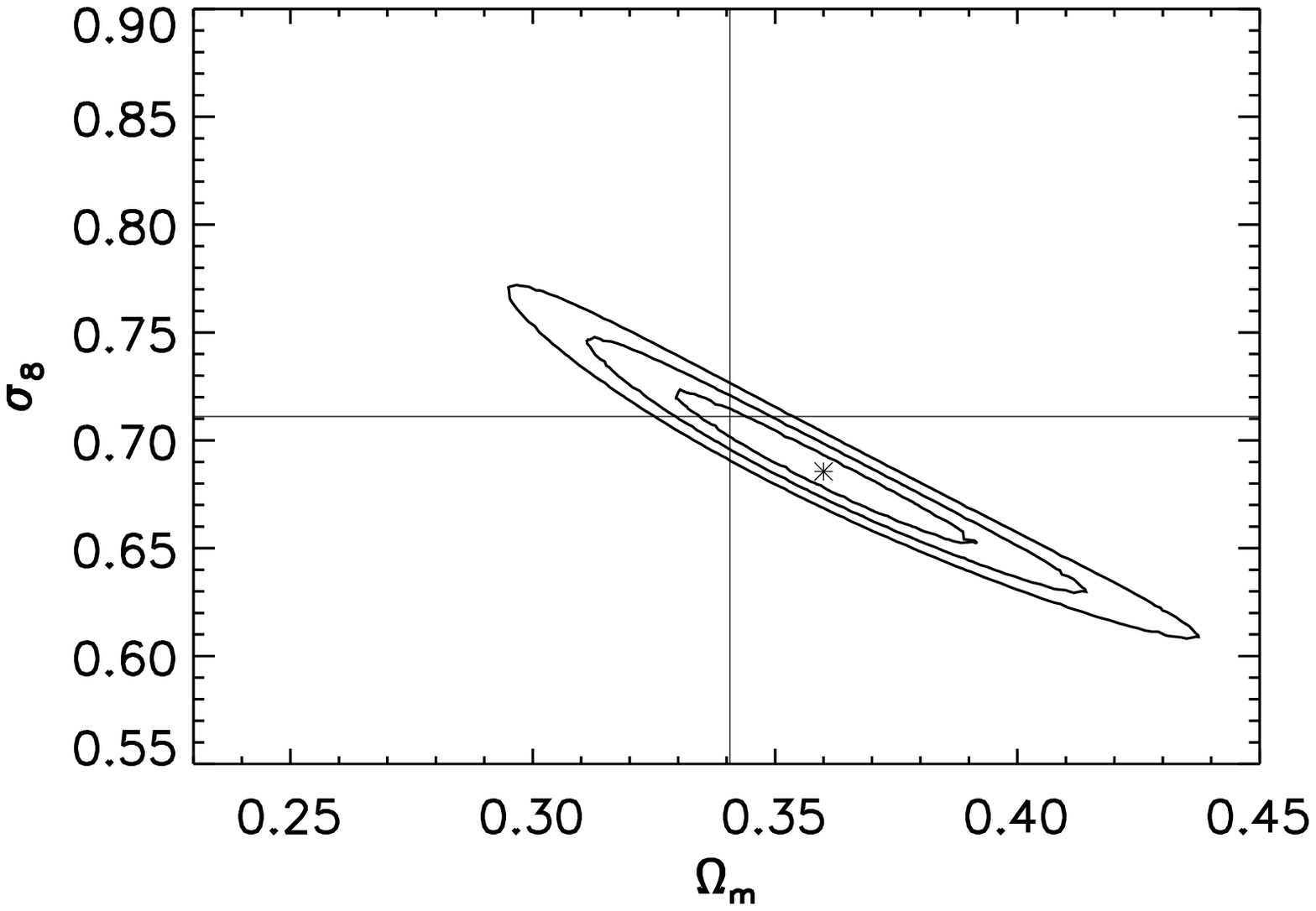,height=5.5cm,width=8.5cm}}
\vspace{-5.5cm}
\centerline{\hspace{ 8.2cm}
\psfig{figure=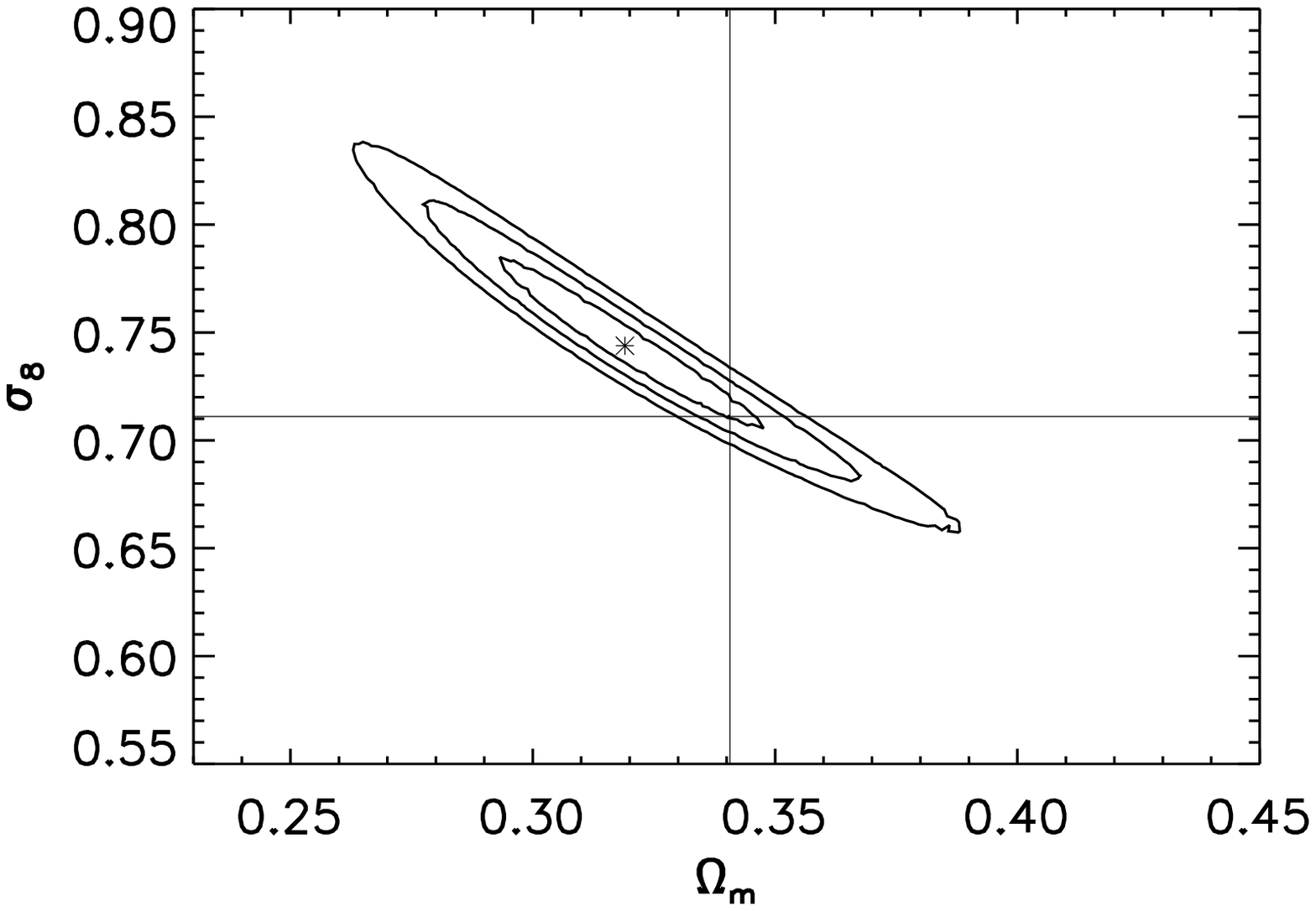,height=5.5cm,width=8.5cm}}
\vspace{-0.35cm}
\caption{\small Likelihood contours ($1$-$3\sigma$ levels for two
degrees of freedom) for models with $h=0.64$ (left) and $h=0.80$
(right). The remaining values of the cosmological parameters and model
assumptions are the same as for the reference KL solution described in
the main text (see also Table\,1). Small stars mark the parameter
values with the highest likelihoods.}
\label{FIG_LOG1}
\end{figure*}

\begin{figure*}
\vspace{-0.0cm}
\centerline{\hspace{-10.5cm}
\psfig{figure=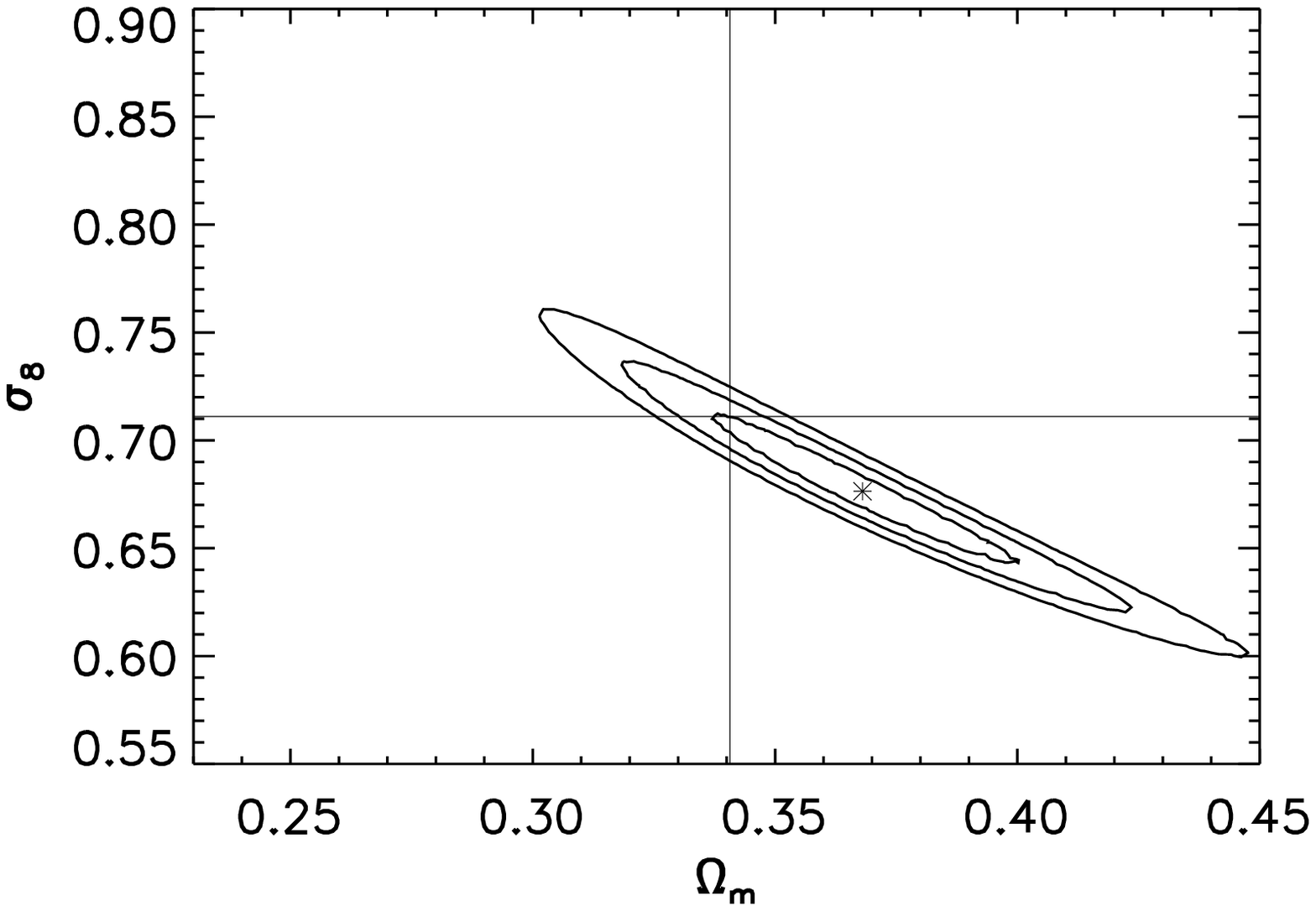,height=5.5cm,width=8.5cm}}
\vspace{-5.5cm}
\centerline{\hspace{ 8.2cm}
\psfig{figure=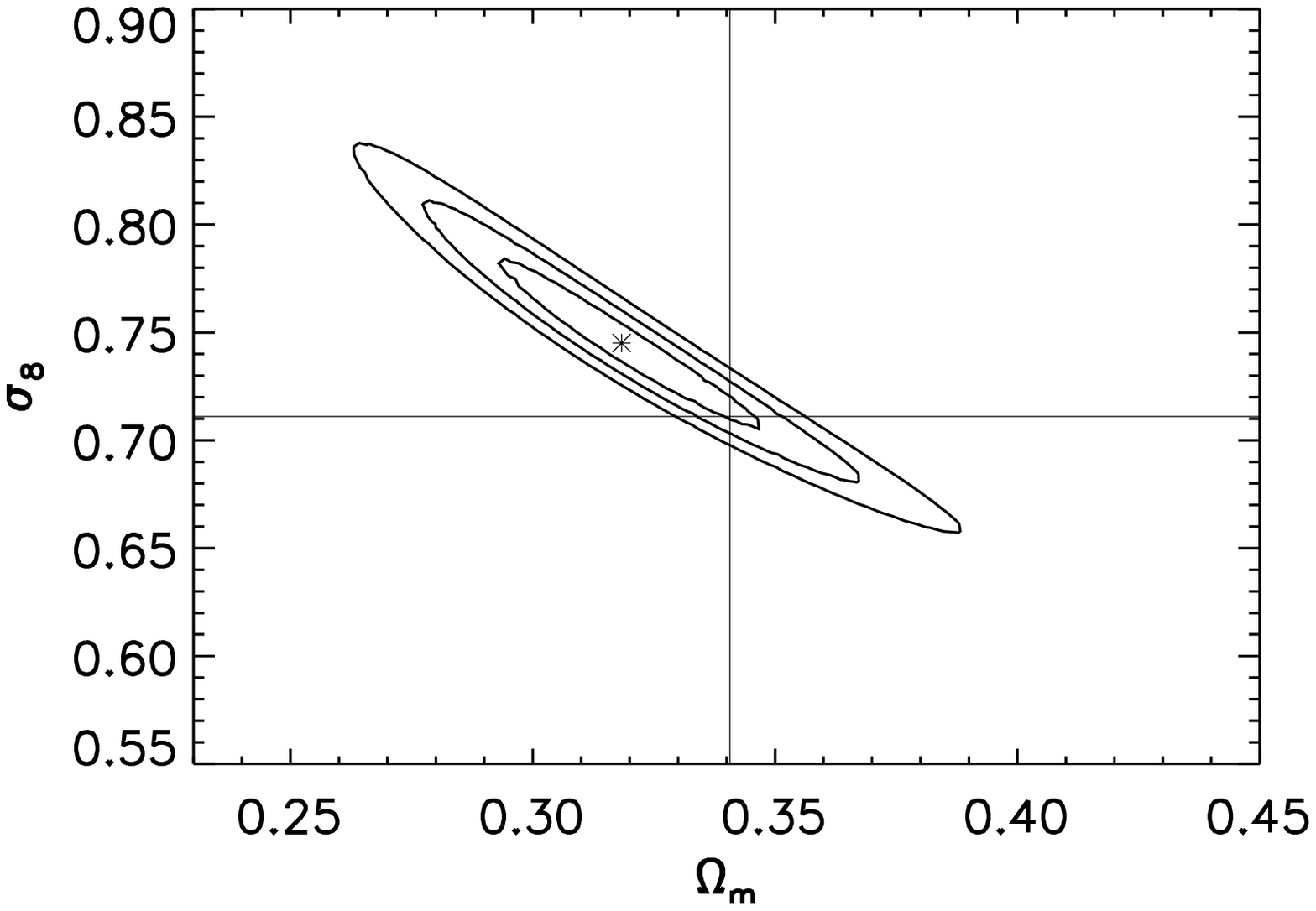,height=5.5cm,width=8.5cm}}
\vspace{-0.35cm}
\caption{\small Likelihood contours for models with $n_{\rm S}=0.9$
(left) and $n_{\rm S}=1.1$ (right). The values of the other parameters
are given in Table\,1.}
\label{FIG_LOG2}
\end{figure*}

\begin{figure*}
\vspace{-0.0cm}
\centerline{\hspace{-10.5cm}
\psfig{figure=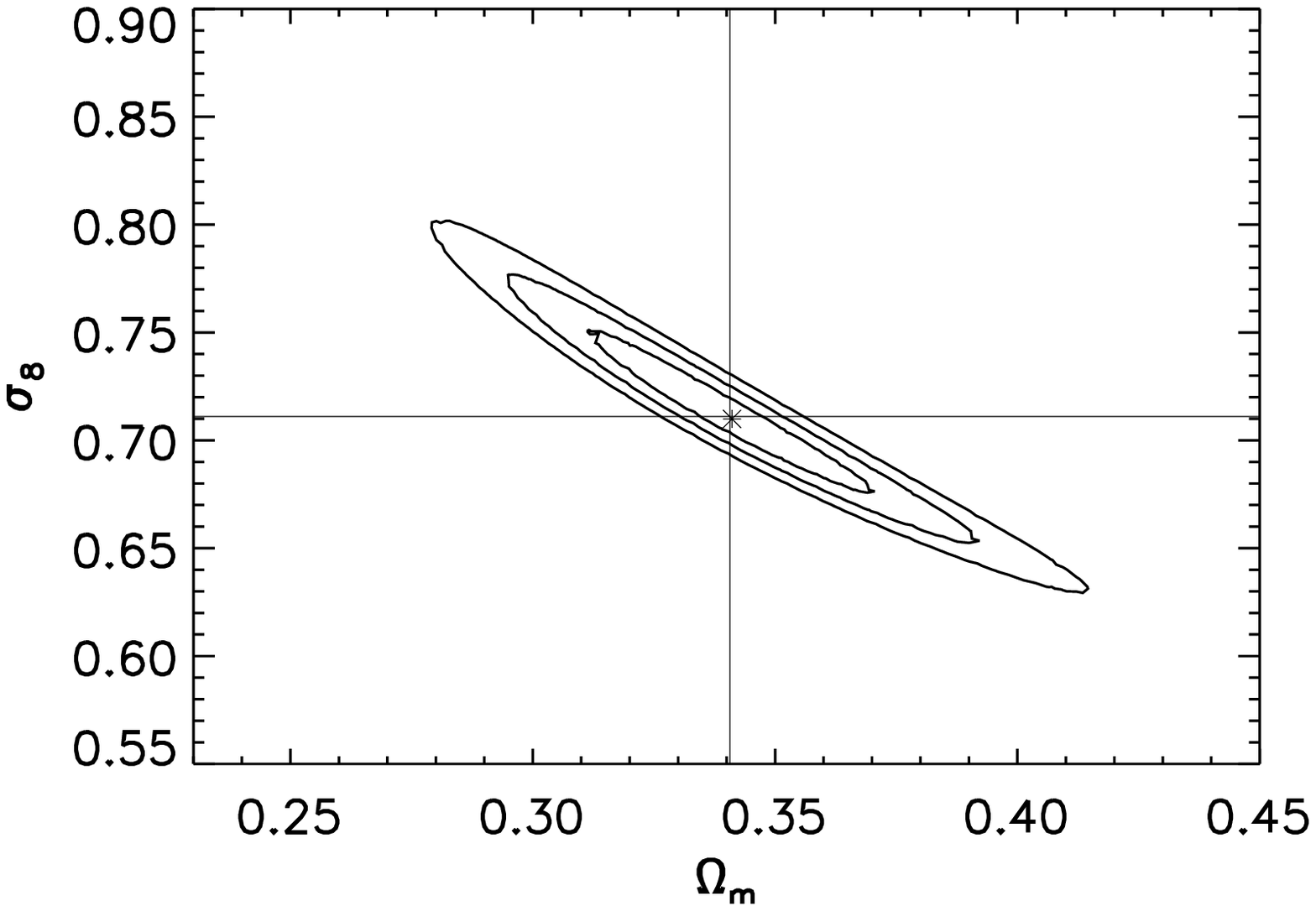,height=5.5cm,width=8.5cm}}
\vspace{-5.5cm}
\centerline{\hspace{ 8.2cm}
\psfig{figure=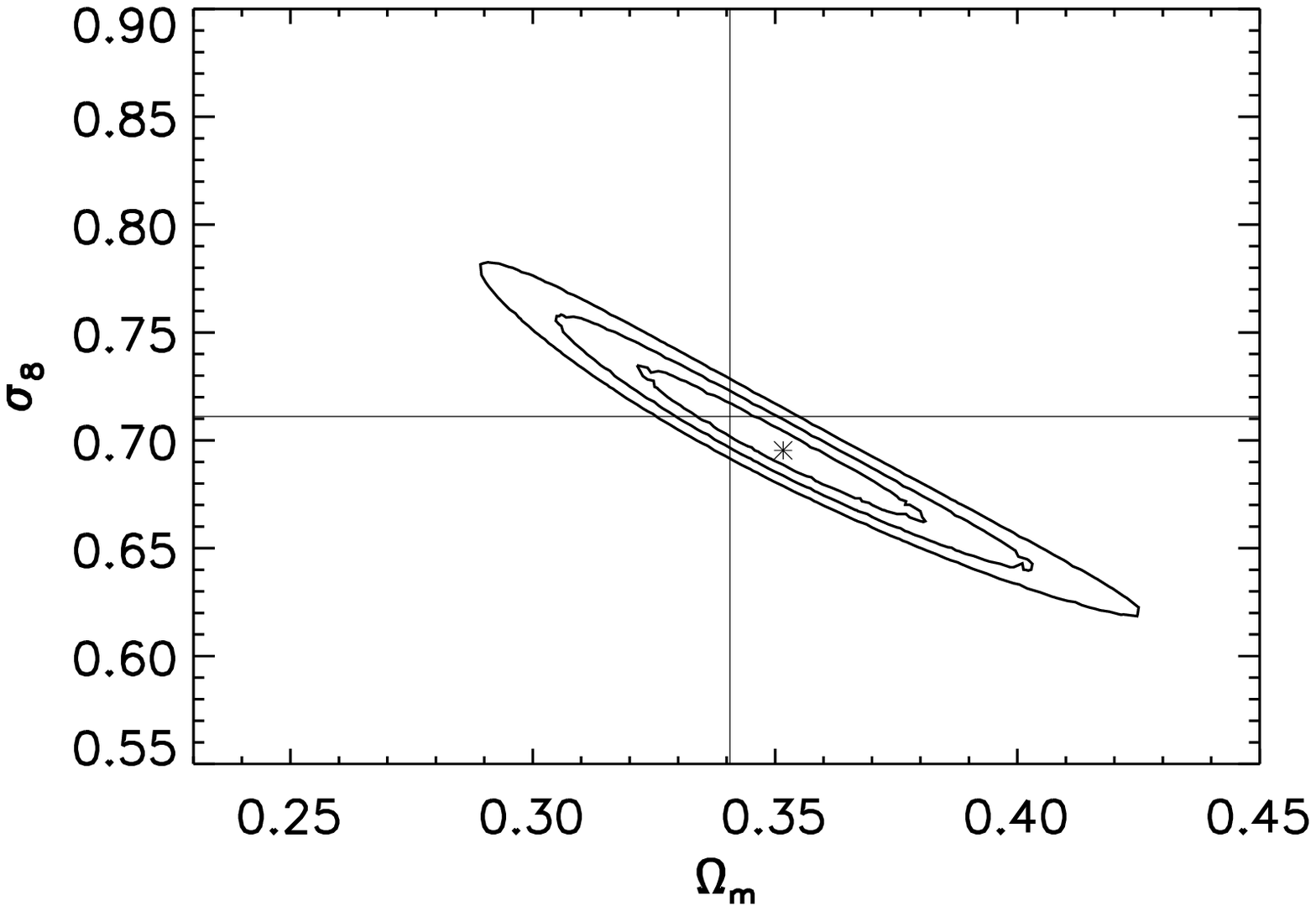,height=5.5cm,width=8.5cm}}
\vspace{-0.35cm}
\caption{\small Likelihood contours for models with
$\Omega_bh^2=0.018$ (left) and $\Omega_bh^2=0.026$ (right). The values
of the other parameters are given in Table\,1.}
\label{FIG_LOG3}
\end{figure*}

\begin{figure*}
\vspace{-0.0cm}
\centerline{\hspace{-10.5cm}
\psfig{figure=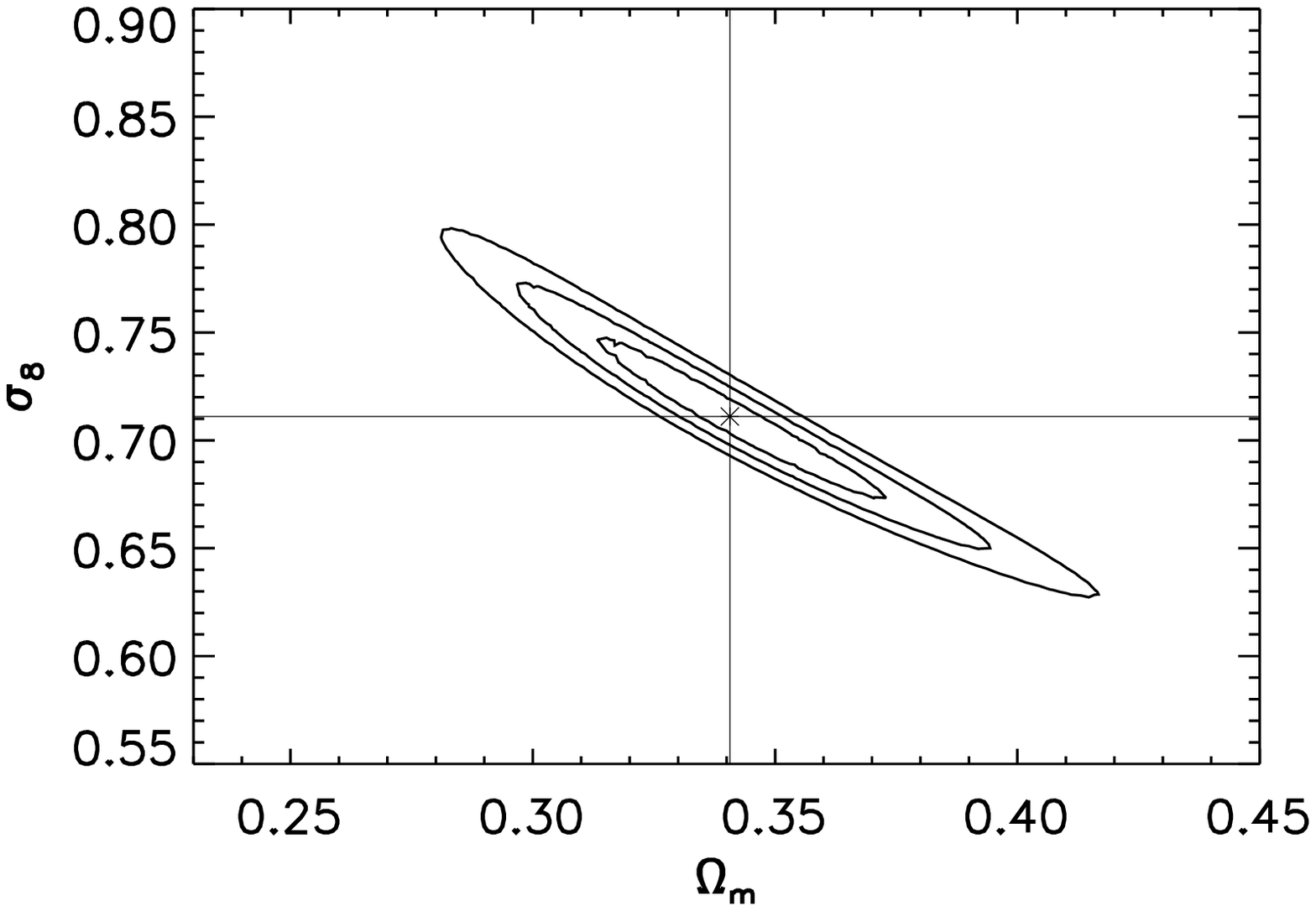,height=5.5cm,width=8.5cm}}
\vspace{-5.5cm}
\centerline{\hspace{ 8.2cm}
\psfig{figure=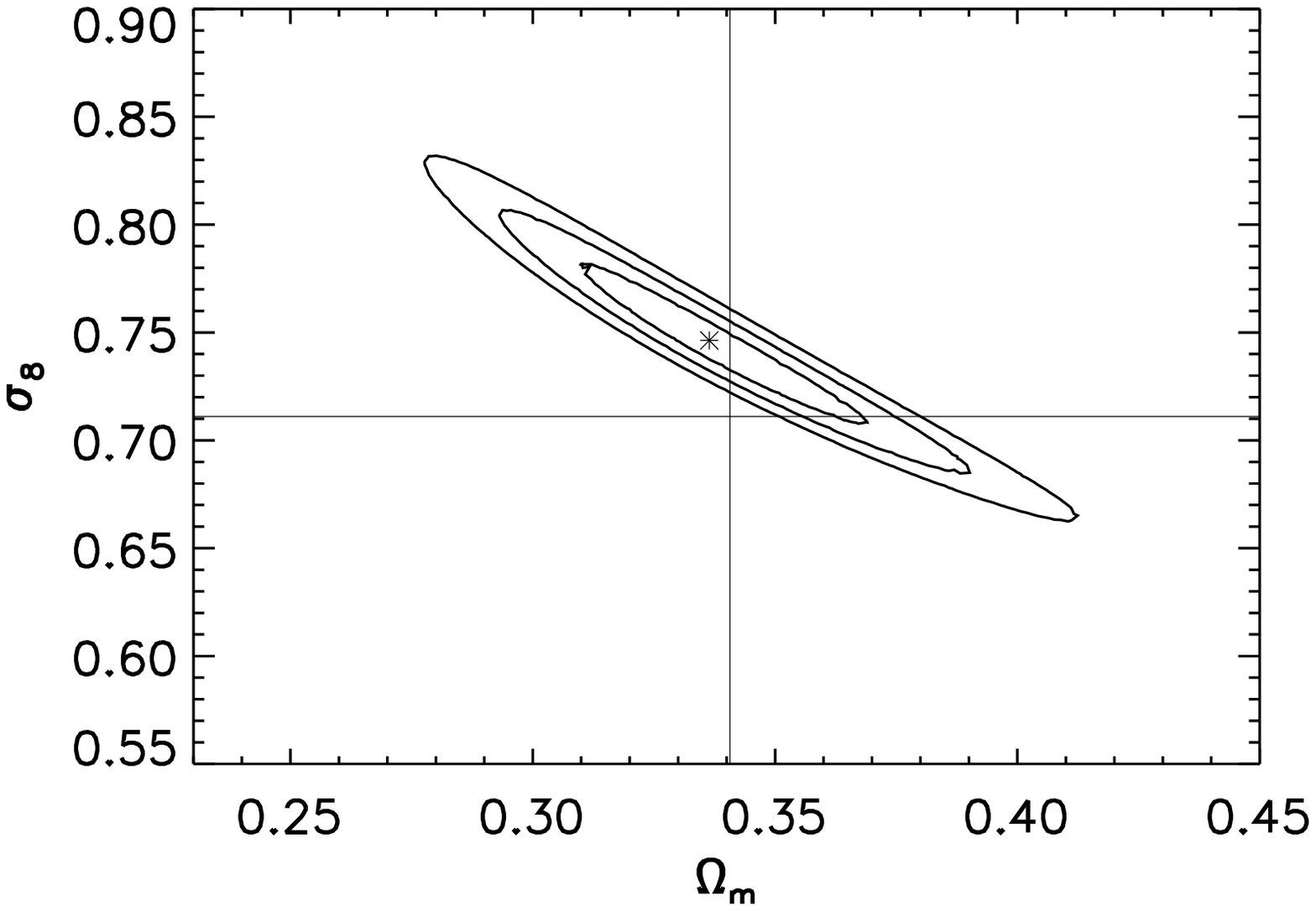,height=5.5cm,width=8.5cm}}
\vspace{-0.35cm}
\caption{\small Likelihood contours for biasing models of Sheth \&
Tormen (1999, left) and of Kaiser (1984, right). The former model
yields the adopted reference solution. The values of the other
parameters are given in Table\,1.}
\label{FIG_LOG4}
\end{figure*}

\begin{figure*}
\vspace{-0.0cm}
\centerline{\hspace{-10.5cm}
\psfig{figure=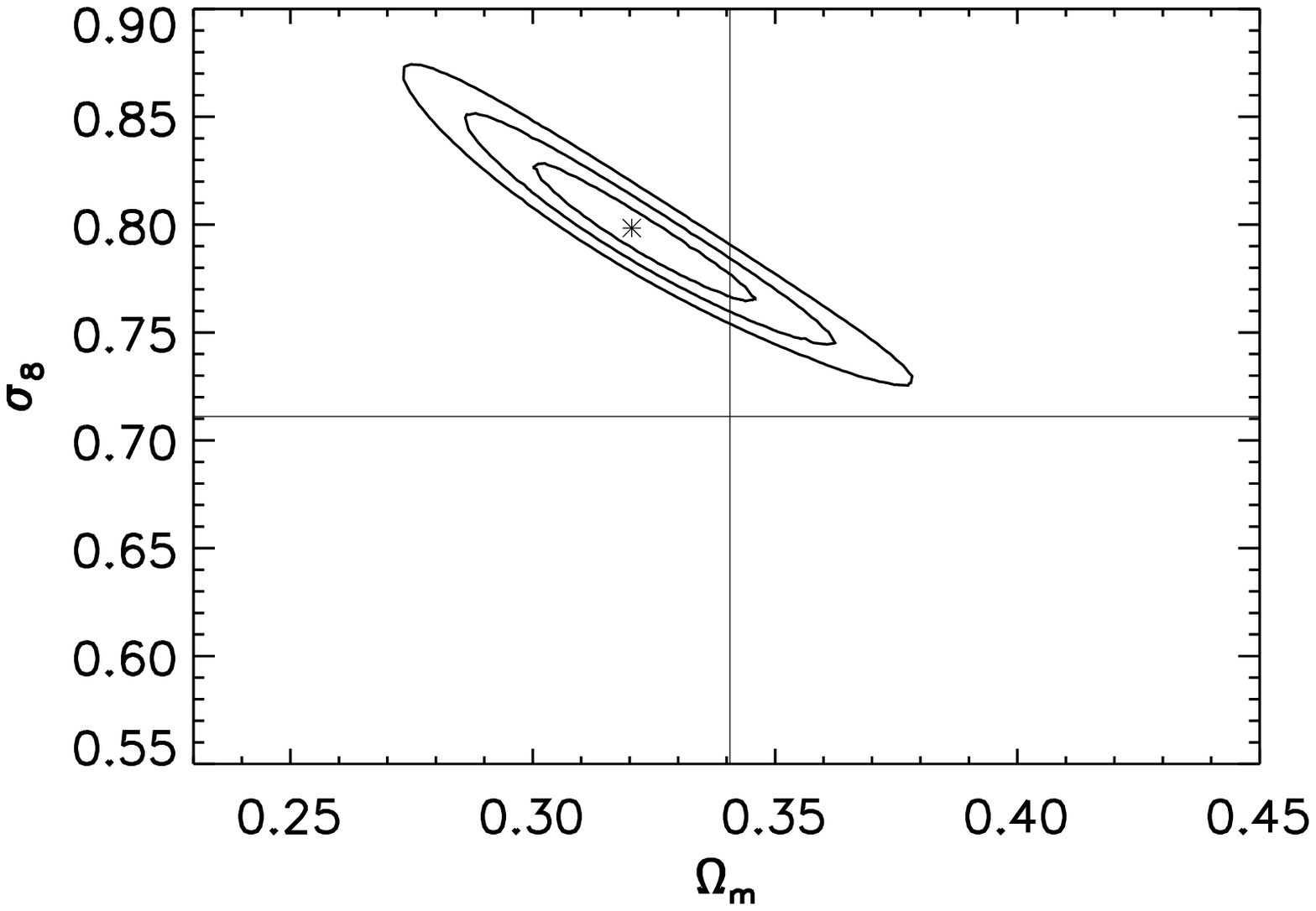,height=5.5cm,width=8.5cm}}
\vspace{-5.5cm}
\centerline{\hspace{ 8.2cm}
\psfig{figure=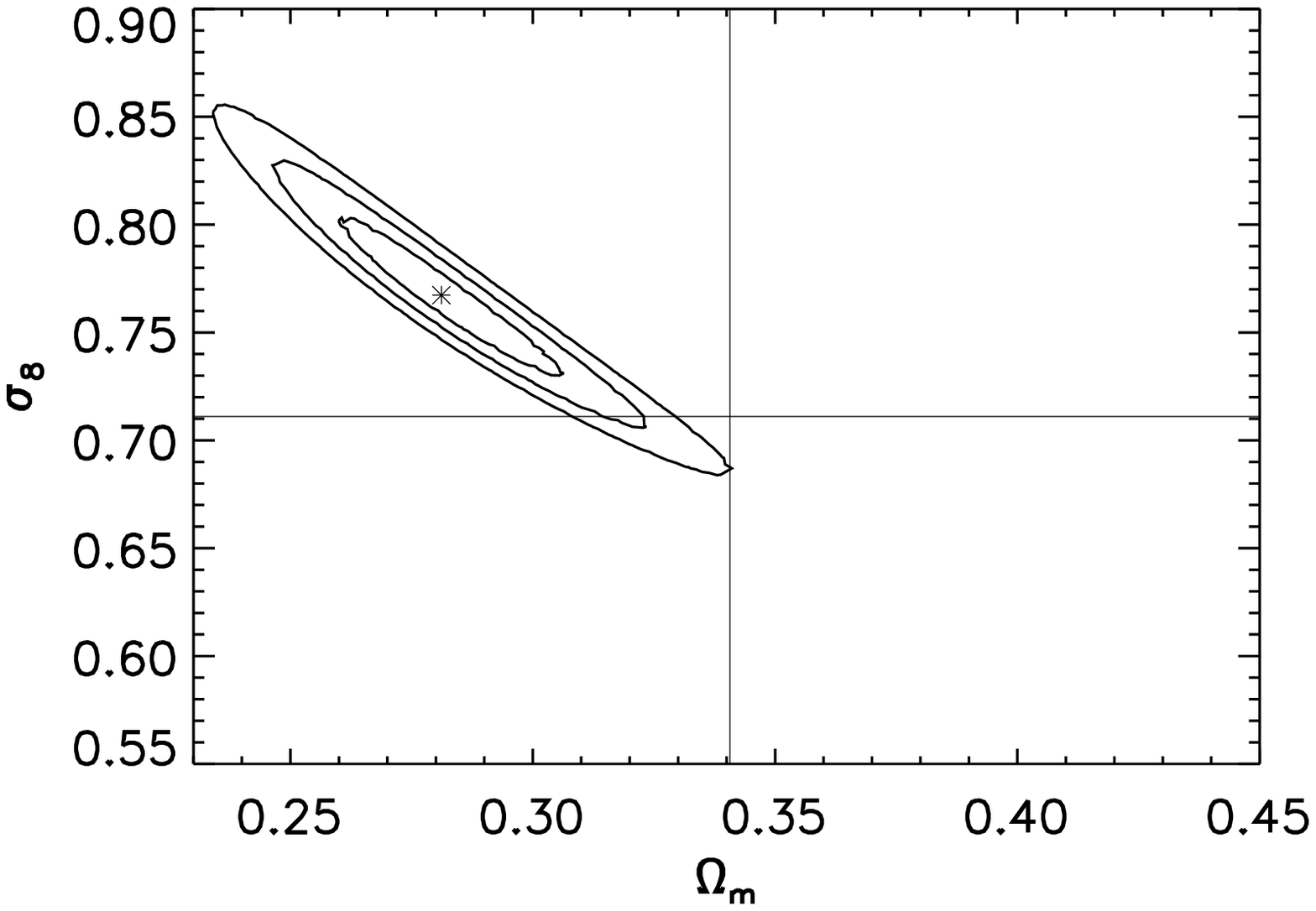,height=5.5cm,width=8.5cm}}
\vspace{-0.35cm}
\caption{\small Likelihood contours for mass/X-ray luminosity
relations with $A=-20.055$, $\alpha=1.652$, $\sigma_{\rm eff}=19\%$
(left) and $A=-18.320$, $\alpha=1.538$, $\sigma_{\rm eff}=25\%$
(right); definition of $A$ and $\alpha$ as in RB02. The values of the
other parameters are given in Table\,1.}
\label{FIG_LOG5}
\end{figure*}

\section{Model specifications}\label{MODEL}

We briefly outline the model for the observed cluster power spectrum
$P_{\rm obs}(k)$ described in more detail in Paper\,II and a new model
for the average cluster abundance $\vec{N}$ which replaces the
empirical approach used in Paper\,II.

The Gaussianity of the KL coefficients $\vec{B}$ measured with the
REFLEX clusters on large scales together with the linearity of the KL
transform suggest that the large-scale cosmic matter distribution is
Gaussian as well and can thus be characterized by a (linear theory)
matter power spectrum $P(k)$. We assume adiabatic scalar perturbations
and a cold dark matter (CDM) plus baryon fluid as suggested by the
observed power spectrum of CMB anisotropies (see recent CMB
measurements mentioned in Sect.\,\ref{INTRO}). The resulting $P(k)$ is
determined by the transfer function $T_x(k)$, the spectral index
$n_{\rm S}$ of the initial scalar fluctuations, and the normalization
parameter $\sigma_8$. The corresponding transfer functions of
Eisenstein \& Hu (1998) are used, providing a more accurate
description of the linear theory power spectrum than fitting equations
of the kind given in Bardeen et al. (1986). Note that the $\sigma_8$
as introduced here reflects the amplitude of the power spectrum
without any non-linear corrections.

Generally, $P_{\rm obs}(k)$ is an average over evolving matter power
spectra and clusters with different values of the biasing parameter
$b$. We summarize this using the prescription of Matarrese et
al. (1997) and Moscardini et al. (2000). Note that the prescription
implies that the observed mass and redshift-averaged cluster power
spectrum has the same shape, i.e., functional form as the underlying
matter power spectrum. In order to illustrate the sensitivity of the
results to the biasing model we use for $b$ alternatively the model of
Sheth \& Tormen (1999) and Kaiser (1984). We verified the model of
$P_{\rm obs}(k)$ (see Paper\,II) with a large set of cluster samples
extracted from the Hubble Volume Simulations (e.g., Evrard et al. 2002
and the references given therein).

For the average cluster abundances the following model is used. For an
unclustered distribution, the average number of clusters expected in a
small cell centered at Right Ascension $\alpha$, Declination $\delta$,
and redshift $z$, per unit redshift and solid angle $\omega$ is
\begin{eqnarray}\label{ML2}
\frac{dN(\alpha,\delta,z)}{dz\,d\omega}\,=\,
\frac{c\,r^2(z)}{H(z)}\,\int_{M_{\rm
lim}(\alpha,\delta,z)}^\infty\,dM\,\frac{dn(M,z)}{dM}\,,
\end{eqnarray}
with the usual notation $H(z)\,=\,H_0\,E(z)$ and
$r(z)\,=\,c/H_0\int_0^zE^{-1}(z')dz'$. The average number of clusters
expected in a KL cell thus is $N_i\,=\,\int_{V_i}dN$. Recent CMB
measurements strongly suggest a flat space (see Sect.\,\ref{INTRO}) so
that $E^2(z)\,=\,\Omega_m(1+z)^3+\Omega_\Lambda$. The mass limit
$M_{\rm lim}$ is obtained from the X-ray luminosity limit $L_{\rm
min}$ at a given $(\alpha,\delta,z)$ and the empirical mass/X-ray
luminosity relation of RB02 including its intrinsic scatter (see
Sect.\,\ref{PRAC}). They measured the masses within a radius with the
average density contrast 200 related to the zero-redshift critical
density $\rho_c$, assuming an Einstein-de Sitter (EdS) geometry.

For cosmological tests a mass function $dn/dM$ is needed which is
form-invariant under different cosmologies. Jenkins et al. (2001)
found from numerical simulations that
\begin{eqnarray}
\frac{dn(M,z)}{dM}\,=\,-0.315\,\frac{\bar{\rho}(z)}{M}\,
\frac{1}{\sigma(M)}\,\times\nonumber
\end{eqnarray}
\begin{equation}\label{ML5}
\hspace{1.25cm}\left|\frac{d\sigma(M)}
{dM}\right|\,\exp\left\{-\left|0.61-\ln\left[\sigma(M,z)\right]\right|^{3.8}\right\}
\end{equation}
obeys this invariance property when the masses $M$ are defined by a
spherical overdensity of $18\pi^2$ related to the background matter
density $\bar{\rho}=\Omega_m\,\rho_c$.  The growth of cosmic
structures introduces a redshift-dependent $\sigma$ in the mass
function, giving the standard deviation of the matter density field
smoothed by a spherical top-hat filter with the radius
$R=[2GM/\Omega_mH^2]^{1/3}$.

In order to combine the empirical mass/X-ray luminosity relation and
the mass function, two transformations of cluster masses are necessary
which are described in Sect.\,\ref{PRAC} and
Appendix\,\ref{MASSTRANS}. Any change of the values of cosmological
parameters thus changes both the observed mass/X-ray luminosity
relation and the theoretical mass function. Note that the cosmology
dependence of the observed mass/X-ray luminosity relation of RB02
results mainly in a change to the model-dependent virial radius,
whereas for a fixed radius the dependence is significantly smaller.

\section{The REFLEX sample and the values of important model 
parameters}\label{PRAC}

The KL analysis is performed with a subsample of 426 clusters of the
452 REFLEX clusters. The clusters of the subsample are selected
according to the following criteria. They are located within comoving
distances $r\le 1.0\,h^{-1}\,{\rm Gpc}$ ($z\le 0.365$). The clusters
have at least 10 X-ray source counts detected in the ROSAT energy band
0.5-2.0\,keV, X-ray luminosities $L_{\rm X}\ge 2.5\times
10^{42}\,h^{-2}\,{\rm erg}\,{\rm s}^{-1}$ and X-ray fluxes $S_{\rm
X}\ge 3.0\times 10^{-12}\,{\rm erg}\,{\rm s}^{-1}\,{\rm cm}^{-2}$ in
the energy band $0.1$-$2.4$\,keV. The clusters are located in an area
of 4.24\,sr in the southern hemisphere with Declination $\le
2.5$\,deg, excluding galactic latitudes $|b|\le20$\,deg and some
additional crowded fields like the Magellanic Clouds. More information
about the sample construction can be found in B\"ohringer et
al. (2001).

Several tests with observed and simulated data suggest that the REFLEX
sample is at least 90\% complete up to the lower X-ray luminosity
limit given above (B\"ohringer et al. 2001, Schuecker et al. 2001). In
the computation of the X-ray luminosities a $\beta$-model based
correction is applied to correct the observed flux for the missing
flux outside the observational apperature radius. The correction is on
average about 10\% and the correction procedure has been tested with
Monte-Carlo simulations (B\"ohringer et al. 2002).  For the cosmic
$K$-corrections we use a refined Raymond-Smith code, where the X-ray
temperatures $T_{\rm X}$ are estimated with the $L_{\rm X}$-$T_{\rm
X}$ relation of Markevitch (1998) without `cooling flow' corrections
(ignoring higher-order effects caused by the fact that the latter
relation was obtained for non-total X-ray luminosities).

For the transformation of the X-ray luminosity limit $L_{\rm min}(z)$
to the corresponding mass limit defined within an EdS geometry (see
Sect.\,\ref{MODEL}) the empirical mass/X-ray luminosity relation of
RB02 is used with the particular parameter values $A=-20.055$ and
$\alpha=1.652$ (following their notation). They measured the masses of
106 galaxy clusters with redshifts generally below $z=0.1$ mainly with
ROSAT PSPC pointed observations and gas temperatures as published
mainly from ASCA observations (assuming spherical symmetry,
isothermality and hydrostatic equilibrium). The sample is large enough
to give a statistically representative summary of the local mass/X-ray
luminosity relation of massive clusters and a rough but useful
estimate of its intrinsic scatter (see below). In order to test the
sensitivity of the cosmological results on the representativity of the
sample used to derive the mass/X-ray luminosity relation, we
alternatively work with an empirical relation derived from a smaller
subsample of 63 clusters of RB02 which is, though strictly
flux-limited, significantly dominated by clusters with cooling flow
signatures.

Having converted the X-ray luminosity limit in the way described
above, the resulting mass limit is transformed with Eq.\,(\ref{V10})
in Appendix\,\ref{MASSTRANS} into a mass system which is consistent
with the actual values of the cosmological parameters. This
transformation is necessary because the average density contrast used
to define the maximum radius for the mass integration depends on the
values of the cosmological parameters (see, e.g., Lahav et al. 1991).
After a further transformation using (Eq.\,\ref{V11}) these masses are
in the system introduced by Jenkins et al. (2001) so that
Eq.\,(\ref{ML2}) can be integrated to give the expected average number
of clusters.

To be more specific, the computations of the two transformations are
based on a general relation between cluster mass and density threshold
(see Eq.\,\ref{V5}). We further assume that the redshift-dependency of
the critical density contrast might introduce a possible
redshift-dependency of the cluster masses virialized at $z$ (see,
e.g., Eq.\,\ref{V10}) which is of the form $M(z)\sim\,M(0)/E(z)$ with
the $E(z)$ function already introduced in Sect.\,\ref{MODEL} (see also
Mohr et al. 2000). Finally, the Jenkins et al. mass function is
defined relative to the $\Omega_m$-dependent average mass density. The
transformation of the masses (still defined relative to the critical
-- though cosmology-dependent -- average density contrast) to the
masses used for the Jenkins et al. mass function is accomplished via
(\ref{V11}). For a given mass density profile (see below) all these
mass transformations can be computed in a straightforward way, and
they have to be re-computed every time when a new cosmological model
is tested in the likelihood analysis.

The integration of the mass function includes a convolution which
takes into account (i) the {\it intrinsic} scatter of the mass/X-ray
luminosity relation (possibly caused by `cooling flows' and cluster
mergers) estimated to be about $\sigma_M=20\%$ in mass (see below),
and (ii) the random flux (luminosity) errors of $\sigma_L=10$-$20\%$
of the REFLEX clusters (B\"ohringer et al. 2001) where the different
error components are assumed to add quadratically. The intrinsic
scatter is computed with the observed total mass scatter of about 50\%
of the empirical mass/X-ray luminosity relation and the individual
mass measurement errors of about 30\% (times a factor 1.5 described
below) obtained with the total sample of 106 clusters of RB02.

However, the mass measurement error of about 30\% is {\it formal} and
assumes, e.g., a constant cluster temperature profile and a symmetric
mass distribution, which is likely not the case. For example, Evrard
et al. (1996) studied the accuracy of X-ray mass estimates using
gasdynamic simulations and found for $\Omega_m=1$ and a critical
density contrast of $250$ a ratio of 1.15 for the isothermal to the
non-isothermal mass estimates and a related $1\sigma$ random mass
error of 36\%. Moreover, the Chandra observation of the elliptical
cluster RBS797 (Jetzer et al. 2002) give differences between the
masses derived under the assumption of spherical and ellipsoidal
cluster shapes ranging from 10\% (oblate) to 17\% (prolate). More
realistic mass errors could thus easily be larger than the given
formal error by a factor of 1.5. Taking this factor into account we
get the above-mentioned maximum estimate of the intrinsic scatter of
$\sigma_M=20$\%.

The effective scatter in mass, $\sigma_{\rm
eff}=\sqrt{\sigma_M^2+(\sigma_L/\alpha)^2}$, includes also the scatter
introduced by the flux errors of the REFLEX clusters. The range of
expected $\sigma_L$ values from 10 and 20\% yields, for
$\sigma_M=20\%$, values between $\sigma_{\rm eff}=21$\% and $23$\%. In
order to estimate the effects related to our incomplete knowledge of
$\sigma_M$ contributing to $\sigma_{\rm eff}$ on the cosmological
constraints, we finally assumed in our tests $\sigma_{\rm eff}$ values
ranging from 19\% to 28\% with a default value of 25\%.

For all mass transformations the Navarro et al. (1997) mass density
profile is used with a redshift and mass-independent concentration
parameter of $c=5$ typical for X-ray clusters of galaxies (see
simulations of Navarro et al. 1997 and, e.g., the observed average
value $c=5.2$ obtained by Allen et al. 2002). The sensitivity of this
specific choice on the final results is tested by using alternatively
$c=4$ and $c=6$. We thus assume that the REFLEX clusters do not show
any significant evolution up to $z=0.3$ as suggested by the
redshift-independent distribution of the comoving number densities of
the REFLEX clusters (Paper\,I, see also Gioia et al. 2001 and Rosati
et al. 2002). Therefore, all mass transformations derived in
Appendix\,\ref{MASSTRANS} and model mass functions are evaluated at
the formal redshift $\bar{z}=0.0$ where the empirical mass/X-ray
luminosity relation was referred to, or alternatively $\bar{z}=0.05$,
that is the mean redshift of the cluster sample used to determine the
relation.

The REFLEX clusters are counted in 360 cells (spherical coordinates):
6 angular bins in Right Ascension, 6 bins in Declination, and 10 bins
along the comoving radial axis. With standard linear algebra codes
(Press et al. 1989) we compute the eigenvectors and eigenvalues of the
whitened correlation matrix $R'$. Tests of the KL method based on 27
independent REFLEX-like cluster samples extracted from the Hubble
Volume Simulation ensure the correct handling of the data (see
Appendix of Paper\,II). The higher order KL modes obtained for the
present investigation show an increasing number of zero-crossings and
a decreasing amplitude (by definition). The same behaviour is found in
our previous KL analysis (see Figs.\,1 and 2 in Paper\,II) and is thus
not illustrated again.

Finally, it should be noted that the model covariance matrix $C$ is
not diagonal unless the fiducial model used to compute the KL
eigenvectors is identical to the model used to compute $R'$. However,
we found that a change to another cosmology (e.g., EdS) has a small
effect on the eigenvectors, slightly broadening the likelihood
contours, but leaving the location of the likelihood maximum almost
unchanged, as first noted in Vogeley \& Szalay (1996). Tests show that
for different initial values of $\Omega_m$ and $\sigma_8$, the KL
solutions converge to similar $\Omega_m$ and $\sigma_8$ values which
we then adopted as our fiducial cosmology (see
Sect.\,\ref{INTRO}). Hence, the fiducial cosmology is close to our
final result (Sect.\,\ref{RESULTS}) so that the effects mentioned
above are negligible.

\begin{figure}
\vspace{-0.0cm}
\centerline{\hspace{0.0cm}
\psfig{figure=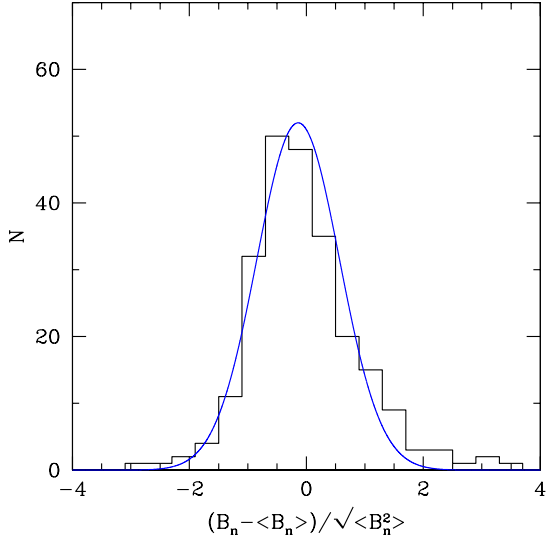,height=8.5cm,width=8.5cm}}
\vspace{-0.7cm}
\caption{\small Histogram of the KL coefficients of the REFLEX cluster
sample obtained with the reference solution (Eq.\,\ref{E1}), and
superposed unit-variance Gaussian profile shifted by $-0.14\sigma$ in
order to get a better representation of the observed data. The
coefficients are normalized by the eigenvalues of the REFLEX sample
correlation matrix.}
\label{FIG_DB}
\end{figure}

\section{Results}\label{RESULTS}

Table\,1 summarizes the basic results for $\Omega_m$ and $\sigma_8$
including their formal $1\sigma$ standard deviations obtained with the
REFLEX cluster sample for different priors under the general
assumption of a non-evolving cluster population and geometrically flat
cosmologies. Typical likelihood contours are shown in
Figs.\,\ref{FIG_LOG1} to \ref{FIG_LOG5}. They illustrate the
sensitivity of the analysis on important cosmological parameters. For
the comparison of different KL solutions we use as reference the
results obtained with $h=0.70$, $\Omega_bh^2=0.020$, $n_{\rm S}=1.0$,
$\bar{z}=0.0$, $\sigma_{\rm eff}=25\,\%$, an empirical mass/X-ray
luminosity relation with the parameters $A=-20.055$, $\alpha=1.652$
(denoted by M106), and the biasing model of Sheth \& Tormen (1999),
yielding (Fig.\,\ref{FIG_LOG4} left and the 8th row in Table\,1)
\begin{equation}\label{E1}
\Omega_m\,=\,0.341\,^{+0.031}_{-0.028}\,,\quad\quad
\sigma_8\,=\,0.711\,^{+0.039}_{-0.031}\,.
\end{equation} 
The $1\sigma$ random errors do not include cosmic variance. Compared
to the results obtained by utilizing the cluster fluctuations only
(Paper\,II), the present random errors are a factor of 2.0 smaller for
$\Omega_m$ and a factor of 8.6 smaller for $\sigma_8$. This
illustrates the importance of including the cluster abundance as a
second criterion because it appears to be the main driver of the high
accuracies reached in the present investigation. However, the results
(\ref{E1}) are more sensitive to systematic errors compared to those
solely obtained with the fluctuation analysis. Therefore, we have to
evaluate the systematic errors introduced by the specific values of
the priors and model assumptions in more detail.

As expected, small systematic differences
($\Delta\Omega_m,\Delta\sigma_8$) are found when the Hubble constant
$h$ (Freedman et al. 2001) and the spectral index of the initial
scalar fluctuations $n_{\rm S}$ (see references given in
Sect.\,\ref{INTRO}) are changed within their respective $1\sigma$
ranges around the priors of the reference solution (see
Figs.\,\ref{FIG_LOG1}, \ref{FIG_LOG2}). They are of the order of the
$1\sigma$ random errors given in (\ref{E1}).

The systematics introduced by the $\Omega_bh^2$ ranges of, e.g., Pryke
et al. (2002) appear comparatively small (see Fig.\,\ref{FIG_LOG3}). A
larger effect is expected when the full range $0.0095\le
\Omega_bh^2\le 0.023$ obtained with standard Big Bang nucleosynthesis
would have been used. However, the measurements of Deuterium in quasar
absorption systems if primordial gives $\Omega_bh^2=0.020\pm0.0015$,
which is quite consistent with recent CMB estimates (see review of
Sarkar 2002) and thus provides a good argument for the smaller
confidence range adopted in the present investigation.

In order to test the stability of the results with respect to the
biasing model two extreme cases are selected (see
Fig.\,\ref{FIG_LOG4}). The model of Sheth \& Tormen (ST, 1999)
although motivated by a Press-Schechter like prescription is
calibrated with dark matter simulations.  The pure statistical biasing
model of Kaiser (KA, 1984) is mainly based on the Gaussianity of the
cosmic matter field. On the large scales studied in the present
investigation we do not expect large differences between the two
models, as supported by the results obtained with the KL analysis (see
Fig.\,\ref{FIG_LOG4} and Table\,1).

\begin{table*}
{\bf Table\,1.} Constraints on $\Omega_m$ and $\sigma_8$ and their
$1\sigma$ errors (without cosmic variance) obtained with the KL method
for the REFLEX cluster sample assuming a flat space. $h$ Hubble
constant. $\Omega_bh^2$: baryon density. $n_{\rm S}$ spectral index of
initial scalar fluctuations. $b$ biasing model: ST Sheth \& Tormen
(1999); KA Kaiser (1984). $\sigma_{\rm eff}$ relative effective
scatter of mass of the empirical mass/X-ray luminosity relation given
in percent. $M/L$ empirical mass luminosity relation: $M/L=106$
corresponding to $A=-20.055$, $\alpha=1.652$ (obtained in RB02 with
106 clusters); $M/L=63$ corresponding to $A=-18.320$, $\alpha=1.538$
(obtained with 63 clusters). $\Omega_m$, $\sigma_8$ and their
$\pm1\sigma$ errors (no cosmic variance); $c$ concentration parameter
of the NFW profile. $\Delta\Omega_m$ and $\Delta\sigma_8$: systematic
differences in $\Omega_m$ and $\sigma_8$ between actual KL solution
and reference solution.\\
\vspace{-0.1cm}
\begin{center}
\begin{tabular}{ccccccc|c|cc|c|ccccc}
$h$ & $n_{\rm S}$ & $\Omega_bh^2$ &  $b$ & $\sigma_{\rm eff}$ & $M/L$ &$c$& $\Omega_m$ &
$+1\sigma$ & $-1\sigma$ & 
$\sigma_8$ & $+1\sigma$ & $-1\sigma$ & $\Delta\Omega_m$ &
$\Delta\sigma_8$\\
\hline
$0.64$ & $1.0$ & $0.020$ & ST & 25\% & 106 &5& $0.360$ & $0.030$ & $0.030$ &
$0.686$ & $0.034$ & $0.036$ & $+0.019$ & $-0.025$ \\
$0.80$ & $1.0$ & $0.020$ & ST & 25\% & 106 &5& $0.319$ & $0.027$ & $0.027$ &
$0.744$ & $0.041$ & $0.034$ & $-0.022$ & $+0.033$ \\
\hline
$0.70$ & $0.9$ & $0.020$ & ST & 25\% & 106 &5& $0.368$ & $0.032$ & $0.031$ &
$0.676$ & $0.036$ & $0.030$ & $+0.027$ & $-0.035$ \\
$0.70$ & $1.1$ & $0.020$ & ST & 25\% & 106 &5& $0.318$ & $0.028$ & $0.025$ &
$0.745$ & $0.040$ & $0.039$ & $-0.023$ & $+0.034$ \\
\hline
$0.70$ & $1.0$ & $0.018$ & ST & 25\% & 106 &5& $0.341$ & $0.029$ & $0.029$ &
$0.711$ & $0.042$ & $0.029$ & $\pm0.000$ & $\pm0.000$ \\
$0.70$ & $1.1$ & $0.026$ & ST & 25\% & 106 &5& $0.352$ & $0.028$ & $0.030$ &
$0.695$ & $0.041$ & $0.035$ & $+0.011$ & $-0.016$ \\
\hline
$0.70$ & $1.0$ & $0.020$ & KA & 25\% & 106 &5& $0.336$ & $0.034$ & $0.026$ &
$0.746$ & $0.034$ & $0.036$ & $-0.005$ & $+0.035$ \\
$0.70$ & $1.0$ & $0.020$ & ST & 25\% & 106 &5& $0.341$ & $0.031$ & $0.028$ &
$0.711$ & $0.039$ & $0.031$ & $\pm0.000$ & $\pm0.000$ \\
$0.70$ & $1.0$ & $0.020$ & ST & 25\% & 63  &5& $0.281$ & $0.025$ & $0.021$ &
$0.767$ & $0.035$ & $0.037$ & $-0.060$ & $+0.056$ \\
\hline 
$0.70$ & $1.0$ & $0.020$ & ST & 19\% & 106 &5& $0.320$ & $0.019$ & $0.026$ &
$0.798$ & $0.030$ & $0.035$ & $-0.021$ & $+0.087$ \\
$0.70$ & $1.0$ & $0.020$ & ST & 28\% & 106 &5& $0.420$ & $0.050$ & $0.050$ &
$0.556$ & $0.054$ & $0.040$ & $+0.079$ & $-0.155
$ \\
\hline 
$0.70$ & $1.0$ & $0.020$ & ST & 25\% & 106 &4& $0.348$ & $0.036$ & $0.020$ &
$0.725$ & $0.029$ & $0.046$ & $+0.007$ & $+0.014$ \\
$0.70$ & $1.0$ & $0.020$ & ST & 25\% & 106 &6& $0.339$ & $0.024$ & $0.034$ &
$0.699$ & $0.041$ & $0.027$ & $-0.002$ & $-0.012$ \\
\hline
\hline
\end{tabular}
\end{center}
\end{table*}

More technically, we measured the systematics introduced by the
assumption that the empirical mass/X-ray luminosity relation of RB02
is determined alternatively at the formal sample mean $\bar{z}=0.05$
and not at $\bar{z}=0.0$ yielding $\Delta\Omega_m=-0.004$ and
$\Delta\sigma_8=+0.007$ (no figure). We further tested the sensitivity
of the cosmological parameters on the effective scatter of the
mass/X-ray luminosity relation by alternatively assuming instead of
$\sigma_{\rm eff}=25\,\%$ the effective scatter of $\sigma_{\rm
eff}=19\,\%$ and $28\,\%$ yielding $\Delta\Omega_m=-0.021$,
$\Delta\sigma_8=+0.087$ (compare Fig.\,\ref{FIG_LOG4} left (25\%) with
Fig.\,\ref{FIG_LOG5} left (19\%)) and $\Delta\Omega_m=+0.079$,
$\Delta\sigma_8=-0.155$, respectively (no figure). We also tested the
sensitivity of the cosmological parameters on the chosen mass/X-ray
luminosity relation using instead of the default relation obtained
with the extended RB02 sample of 106 clusters (denoted M106:
$A=-20.055$, $\alpha=1.652$) a relation obtained with a strict
flux-limited sample of 63 clusters (denoted M63: $A=-18.320$,
$\alpha=1.538$) yielding $\Delta\Omega_m=-0.060$ and
$\Delta\sigma_8=+0.056$ (compare Fig.\,\ref{FIG_LOG4} left (M106) with
Fig.\,\ref{FIG_LOG5} right (M63)). Finally, we tested the effect
of a value of the concentration parameter of the NFW mass density
profile different from the default value $c=5$. For $c=4$ and $c=6$ we
obtained $\Delta\Omega_m=+0.007$, $\Delta\sigma_8=+0.014$ and
$\Delta\Omega_m=-0.002$, $\Delta\sigma_8=-0.012$, respectively (no
figure).

For the correct determination of the confidence ranges caused by the
combined effect of all uncertainties of the priors and model
assumptions one has to analyse the complete 8-dimensional parameter
space. The KL method is, however, quite computer-intensive so that the
errors could only be evaluated in the following simplified manner. The
approximate combined effect (conservative upper limit) is estimated by
summing up the individual systematic errors. We obtain the maximum
systematic errors of $\Delta\Omega_m=^{+0.143}_{-0.137}$ and
$\Delta\sigma_8=^{+0.266}_{-0.243}$. A correlation of the systematics
evaluated above is, however, quite unlikely. More realistic error
estimates assume that the systematics are uncorrelated and that the
squared $\Delta$'s can be regarded as the variances of error
distributions assumed to be Gaussian. In this case the errors add
quadratically and we obtain the final systematic errors
\begin{equation}\label{ERR}
\sigma_{\Omega_m}\,=\,^{+0.087}_{-0.071}\,,\quad\quad
\sigma_{\sigma_8}\,=\,^{+0.120}_{-0.162}\,.
\end{equation}
Finally, we tested whether the Gaussianity assumed throughout the
likelihood analysis is supported by the present measurements. Although
the histogram of the normalized KL coefficients (Fig.\,\ref{FIG_DB})
shows a small systematic shift of $0.14\sigma$ and some excess at
large positive fluctuations relative to a zero-mean, unit-variance
Gaussian profile, the latter distribution and the un-shifted observed
distribution must be regarded as selected from the same parent
distribution when tested on the $3\sigma$ confidence level
(KS-test). The present measurements thus support the fundamental
assumption of Gaussianity, although less significant compared to our
first findings (see Paper\,II).

\begin{figure}
\vspace{-0.3cm}
\centerline{\hspace{-1.0cm}
\psfig{figure=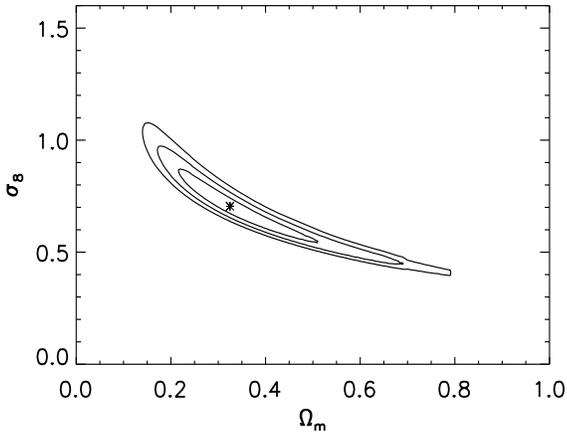,height=6.5cm,width=8.5cm}}
\vspace{-0.25cm}
\caption{\small Likelihood contours obtained with the REFLEX cluster
abundances only (center at $\Omega_m=0.32$ and $\sigma_8=0.71$). The
comparison with Fig.\,\ref{FIG_LOG4} (left) obtained by using both
cluster abundance and fluctuations shows that the KL method breaks the
degeneracy between $\sigma_8$ and $\Omega_m$. Note the different
scales of the diagrams.}
\label{FIG_ABUN}
\end{figure}

\section{Discussion and conclusions}\label{CONCLUSION}

In the present investigation cluster number counts and their spatial
fluctuations are analysed for the first time simultaneously in order
to constrain cosmological models. The test utilizes the
complementarity of clustering and abundance of galaxy clusters and
allows us to fully exploit the cosmological potential of the REFLEX
survey of X-ray clusters (but see below).

For spatially flat cosmologies as suggested by all recent CMB
measurements (Sect.\,\ref{INTRO}) and for a non-evolving cluster
population as suggested by the $z$-independent comoving REFLEX cluster
number densities (Paper\,I), we obtain $\Omega_m=0.341$ for the cosmic
matter density and $\sigma_8=0.711$ for the normalization of the
matter power spectrum. The random errors range between 8-9\% for
$\Omega_m$ and 4-5\% for $\sigma_8$, not including cosmic
variance. Systematic errors can be identified and quantified so that
the whole process appears to be well-controlled. Under the assumption
that the systematics add quadratically the total systematic errors are
found to be about 2.5 and four times larger than the random errors of
$\Omega_m$ and $\sigma_8$, respectively.

Compared to the results obtained with the fluctuations only
(Paper\,II), the combination of abundance and fluctuation measurements
reduces the random errors of $\Omega_m$ by a factor of about two and
of $\sigma_8$ by a factor of about nine. The main driver of the
accuracy of the present results thus appears to be the cluster
abundance, especially for $\sigma_8$ (but see below).

In contrast to many other cosmological parameter estimations we tested
the assumed functional form of the sample likelihood function. As in
Paper\,II we could verify Gaussianity on large scales -- a fundamental
property of the cosmic matter field. In this respect the present
analysis appears to be fully consistent and self-contained.

\begin{figure*}
\vspace{-0.0cm}
\centerline{\hspace{-9.5cm}
\psfig{figure=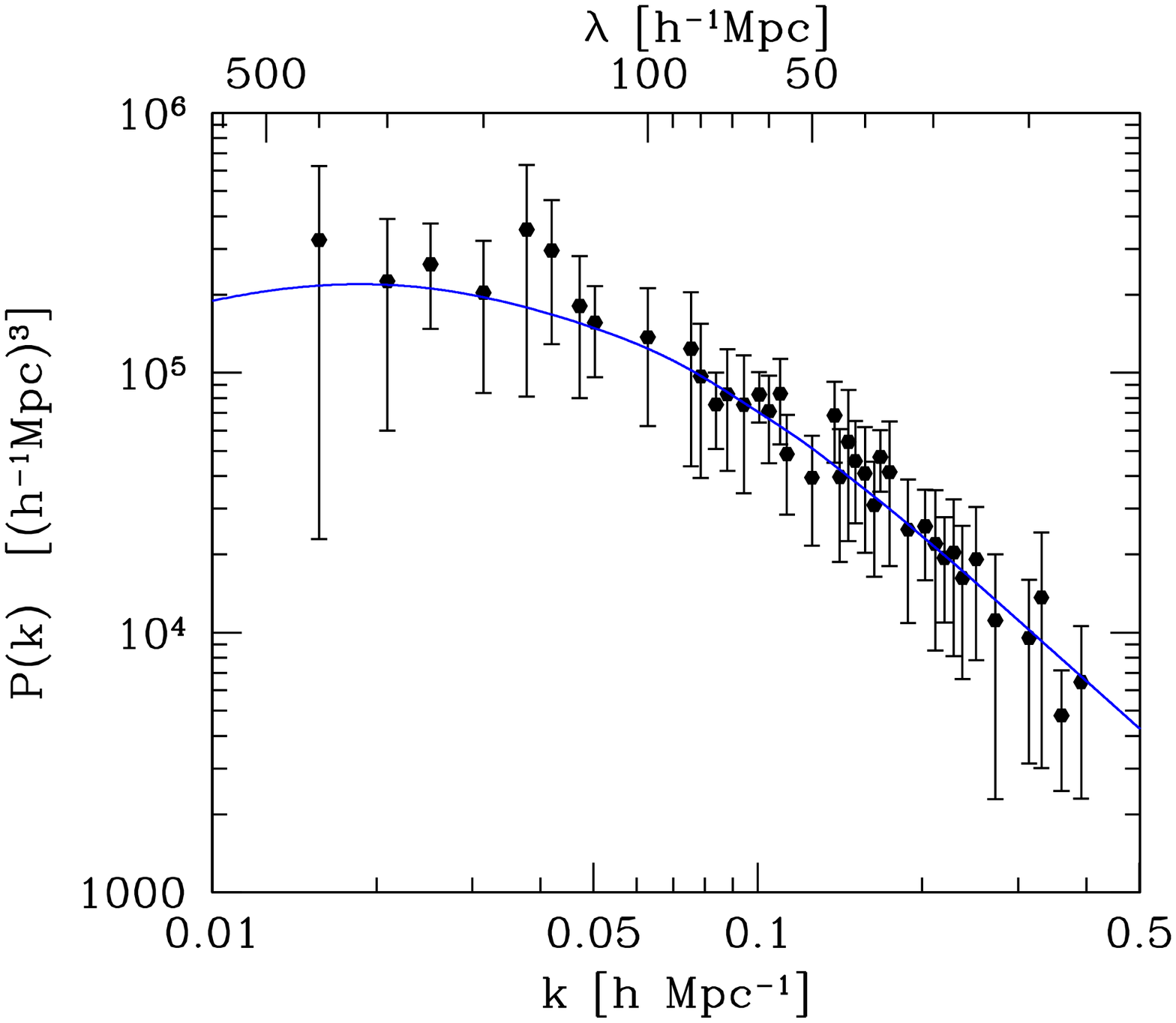,height=8.5cm,width=8.5cm}}
\vspace{-8.5cm}
\centerline{\hspace{ 8.2cm}
\psfig{figure=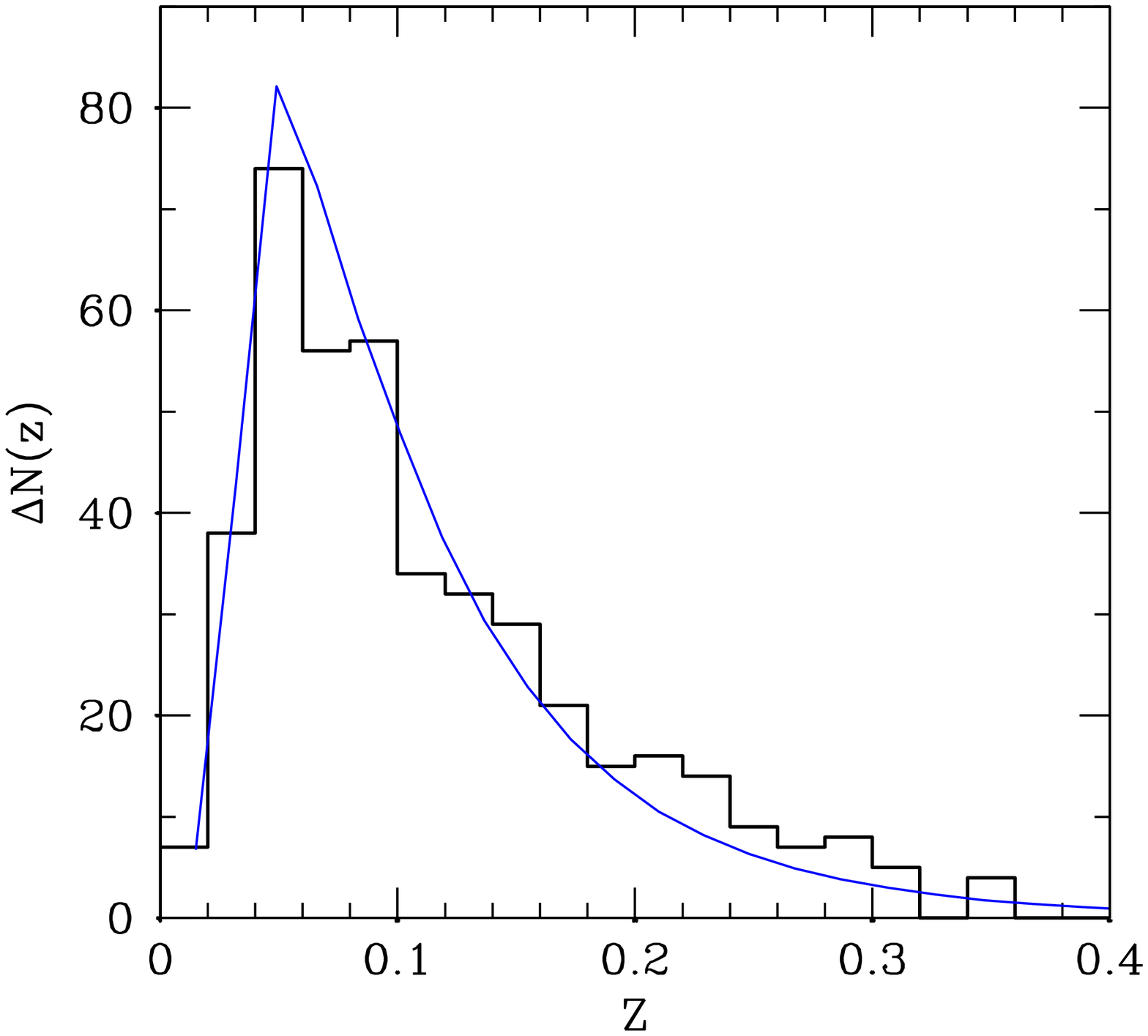,height=8.5cm,width=8.5cm}}
\vspace{-0.65cm}
\caption{\small Comparison of the reference KL solution ($\Omega_{\rm
tot}=1$, $\Omega_m=0.341$, $\Omega_bh^2=0.020$, $h=0.70$, $n_{\rm
S}=1.0$, $\sigma_8=0.711$, $\sigma_{\rm eff}=25\%$, ST biasing) with
REFLEX observations. {\bf Left:} Power spectrum obtained with the
flux-limited REFLEX sample (Paper\,I, points with $1\sigma$ error bars
including cosmic variance) and the KL solution (continuous line). The
theoretical model takes into account the effects of the different
volumes covered by the present KL and the former Fourier analysis,
includes the effects of the baryons, and is transformed into redshift
space using the nonlinear model described in Paper\,I. In order to
make the measured power spectrum less crowded adjacent power spectral
densities and their errors were averaged. {\bf Right:} Redshift
histogram of the REFLEX subsample used here (steps) and the KL
solution (continuous line). }
\label{FIG_COMP}
\end{figure*}

The KL likelihood contours (Figs.\,\ref{FIG_LOG1} to \ref{FIG_LOG5})
show some correlations between $\Omega_m$ and $\sigma_8$, probably
indicating that the primordial part of the power spectrum is still
less constrained by the REFLEX data. The correlations are, however,
restricted to comparatively small $(\Omega_m,\sigma_8)$ ranges so that
it does not make much sense to determine an $\Omega_m$-$\sigma_8$
relation from the data. Therefore, the degeneracy between $\Omega_m$
and $\sigma_8$ can be regarded as broken. The breaking of the
degeneracy by the KL method using both cluster abundance and
fluctuations is nicely seen when the KL reference results
(Fig.\,\ref{FIG_LOG4}, left) are compared with the likelihood contours
obtained with the REFLEX cluster abundance only
(Fig.\,\ref{FIG_ABUN}).

Generally, likelihood analyses which optimize simultaneously two or
more almost independent observational quantities like the power
spectrum and the cluster abundance, try to find the best compromise
between the individual observables and not necessarily the best fit of
the different components. However, Fig.\,\,\ref{FIG_COMP} shows that
the best KL solution obtained with the REFLEX data is consistent with
both the redshift histogram and the redshift-space power spectrum
obtained with a standard Fourier method (Paper\,I). Note that the
power spectrum was measured within a comparatively small volume to
reduce correlations between different fluctuation modes. A further
argument supporting the reliability of the present cosmological
constraints is given by numerical fits of the amplitude of the
observed power spectrum of three volume-limited REFLEX subsamples
using $\Lambda$CDM models which yield $\sigma_8=0.70$ (Paper\,I). This
result is at variance to the standard normalization of the
$\Lambda$CDM model, but in very good agreement with the results of the
present KL analysis.

The value of the cosmic matter density obtained with REFLEX is in good
agreement with other recent independent measurements (see also Turner
2002). For example, Pryke et al. (2002) measured with DASI a matter
density of $\Omega_m=0.40\pm0.15$ for $h=0.72\pm0.08$ and Sievers et
al. (2002) measured with CBI $\Omega_m=0.62\pm0.22$ (weak priors).
Netterfield et al. (2002) obtained for flat geometries
$\Omega_m=0.33\pm0.05$, combining COBE-DMR, BOOMERANG, large-scale
structure, and supernovae type Ia data.  Sievers et al. (2002) found
from the combination of COBE, BOOMERANG, MAXIMA-1, DASI, CBI,
large-scale structure surveys, Hubble parameter determinations, and
supernovae type Ia the value $\Omega_m=0.32\pm0.06$. With the galaxy
power spectrum of the 2dF Galaxy Redshift Survey and a new compilation
of CMB data (COBE, BOOMERANG, MAXIMA-1, DASI, VSA see Scott et
al. 2002, CBI see Mason et al. 2002), Percival et al. (2002) obtained
for a flat cosmology $\Omega_m=0.313\pm0.055$ (all $1\sigma$ errors
without cosmic variance). Therefore, a significant number of
independent applications appear to converge to a low matter density.

The linear theory $\sigma_8$ obtained with REFLEX is lower than the
COBE normalization and most estimates from cluster abundance as
summarized in, e.g., Table\,6 in RB02. Note that values of the
effective scatter $\sigma_{\rm eff}$ of the mass/X-ray luminosity
relation larger than assumed here would increase this descrepancy even
more. However, small $\sigma_8$ values were also obtained by
Markevitch (1998) using a local cluster sample with ASCA X-ray
temperatures and ROSAT X-ray luminosities, by Borgani et al. (2001)
using the ROSAT Deep Cluster Survey, by Seljak (2001) from the
empirical mass/X-ray temperature relation of nearby clusters, by
Viana, Nichol \& Liddle (2002) using SDSS/RASS data and the X-ray
luminosity function of REFLEX clusters of galaxies, and by Bahcall et
al. (2002) using clusters selected from SDSS data. In this context it
is also interesting to note that the KL results are in very good
agreement with the new constraints obtained by Lahav et al. (2002)
from their simultaneous analysis of the amplitudes of the fluctuations
probed by the 2dF Galaxy Redshift Survey and COBE, BOOMERANG,
MAXIMA-1, DASI data, and with the constraints obtained with the KL
analysis of the early Sloan Digital Sky Survey by Szalay et al. (2001)
obtained under similar priors (as described in Lahav et al.). There is
thus increasing observational evidence for a $\sigma_8$ value lower
than previously, although some recent weak lensing results show a
tendency for a higher $\sigma_8$ for given $\Omega_m$ (e.g., Hoekstra
et al. 2002, van Waerbeke et al. 2002). Small $\sigma_8$ values help
(James S. Bullock, private communication) to reduce the apparent
problem of the $\Lambda$CDM model with the standard normalization to
overpredict the number of low-mass satellite galaxies (Klypin et
al. 1999, Moore et al. 1999).

In all the cosmological studies mentioned above including the present,
various model corrections might change the final results by 10\% to
30\% so that more detailed analyses of possible sources of systematic
errors are needed to get below this error level. Our tests for
systematic errors show that the largest systematic errors are
introduced by the mass/X-ray luminosity relation (as expected). In
order to use the full potential of REFLEX-like cluster surveys it is
thus important to have more precise measurement of this function,
i.e., its shape, intrinsic scatter, redshift-dependency, etc. which
are expected to be provided by detailed observations with the Chandra
and XMM satellites. Future REFLEX papers will test carefully the
residual effects of possible evolution of nearby clusters, the
cosmological constant and related quantities (H. B\"ohringer et al.,
P. Schuecker et al., in preparation).

Regardless of these interim restrictions, the present investigation
illustrates that large and fair samples of X-ray clusters of galaxies
give quite clean measurements of the cosmological parameters. In
addition to the abovementioned quality of $\sigma_8$ constraints
traditionally obtained with clusters of galaxies, the $1\sigma$ random
error for the matter density of $\sigma_{\Omega_m}=0.030$ obtained
with the REFLEX data is already close to the random error of
$\sigma_{\Omega_m}=0.010$ expected to be attainable with the CMB
Planck Surveyor satellite and similar to $\sigma_{\Omega_m}=0.015$
attainable with the SNfactory plus SNAP supernovae satellite projects
(see, e.g., Hannestad \& M\"ortsell 2002). Moreover, the cosmological
constraints obtained with cluster data have degeneracies different
from high-$z$ supernovae and CMB anisotropies (Holder, Haiman \& Mohr
2001). Therefore, galaxy clusters can play a significant r\^{o}le in
high precision measurements of cosmological parameters.

\begin{acknowledgements}                                                        
We would like to thank the ROSAT and REFLEX team for their help in the
preparation of the X-ray cluster sample, T.H. Reiprich for very
helpful comments concerning the empirical mass/X-ray luminosity
relation and for critical reading of the manuscript, and the anonymous
referee for some useful suggestions. P.S. acknowledges support under
the grant No.\,50\,OR\,9708\,35.
\end{acknowledgements}

\appendix

\section{Mass transformations}\label{MASSTRANS}

In this appendix we discuss mass transformations which are necessary
when masses $M_\Delta$ are defined within radii $r_\Delta$ which
correspond to different average density contrasts $\Delta$ and when
average contrasts are related to different background densities. We
are particularily interested in the relation between the masses used
in the empirical mass/X-ray luminosity relation of Reiprich \&
B\"ohringer (2002) and the masses used in the theoretical mass
function of Jenkins et al. (2001). To motivate the mass
transformations we first assume an isothermal sphere at redshift $z=0$
and define a virial radius by
\begin{equation}\label{V1}
r_{\rm VIR}\,=\,\left[\frac{M_{\rm
VIR}}{(4\pi/3)\Delta_c\,\rho_c}\right]^{1/3}\,,
\end{equation}
with the average density contrast $\Delta_c$ which is $18\pi^2$ for
the EdS case, and the critical EdS density $\rho_c=3H_0^2/8\pi G$. The
virial relation then reads $G\,M_{\rm VIR}\,=\,a\,T_{\rm X}\,r_{\rm
VIR}$, with $T_{\rm X}$ the temperature of the X-ray emitting gas and
$a$ a structural constant. If massive clusters are simply re-scaled
versions of low mass clusters (e.g., Neumann \& Arnaud 2001) then $a$
is a universal constant and one obtains
\begin{equation}\label{V3}
M_{\rm
VIR}\,=\,\left(\frac{a}{G}\right)^{3/2}\frac{1}{\sqrt{4\pi/3}}\,
\frac{1}{\sqrt{\Delta_c}}\frac{1}{\sqrt{\rho_c}}\,\,T_{\rm X}^{3/2}\,.
\end{equation}
The assumption of an isothermal sphere mass distribution thus yields
for a given redshift and temperature the mass-density contrast
relation $M_{\rm VIR}/M_\Delta=(\Delta/\Delta_c)^{0.5}$. Deviations
from isothermality changes this relation as can be seen in Horner et
al. (1999) who found from spatially resolved temperature measurements
the empirical relation $M_{\rm
VIR}/M_\Delta=(\Delta/\Delta_c)^{0.266}$, consistent with the mass
density profile $\rho(r)\sim r^{-2.4}$. In order to be consistent with
our reference to high-quality dark matter simulations we assume a
Navarro, Frenk \& White (NFW, 1997) profile. In this case we have to
deal with a mass ratio $M_{\rm
VIR}/M_\Delta\equiv\Pi(\Delta_c,\Delta)$ which can only be obtained
numerically using
\begin{equation}\label{V5}
\frac{M_{\Delta_1}}{M_{\Delta_2}}\,=\,\frac{\Delta_1}{\Delta_2}\,
\cdot\,\left(\frac{r_{\Delta_1}}{r_{\Delta_2}}\right)^3\,,
\end{equation}
\begin{equation}\label{V6}
\frac{M_{\Delta_1}}{M_{\Delta_2}}\,=\,
\frac{\ln\left(1+c\,r_{\Delta_1}/r_{200}\right)\,-\,
\frac{cr_{\Delta_1}/r_{200}}{1+c\,r_{\Delta_1}/r_{200}}}
     {\ln\left(1+c\,r_{\Delta_2}/r_{200}\right)\,-\,
\frac{cr_{\Delta_2}/r_{200}}{1+c\,r_{\Delta_2}/r_{200}}}\,,
\end{equation}
where $c$ is the concentration parameter which we assume to be
independent of mass and redshift. Iterative solution for
$r_{\Delta_1}/r_{\Delta_2}$ yields the mass ratio $\Pi$ where the
normalization radius $r_{200}$ of the NFW profile cancels out. Due to
its definition via a mass ratio, $\Pi$ has the property
$\Pi(\Delta_1,\Delta_2)\cdot\Pi(\Delta_2,\Delta_3)=
\Pi(\Delta_1,\Delta_3)$, yielding
\begin{equation}\label{V8}
\frac{M_{\rm VIR}(0)}{M_{200}^{\rm
EdS}(0)}\,=\,\Pi\left(\Delta_c(0),\Delta_c^{\rm
EdS}(0)=18\pi^2\right)\,.
\end{equation}
The arguments indicate that the quantities are determined at redshift
$z=0$. Note that the density contrast $\Delta_c^{\rm EdS}(0)$
approximately defines the same mass $M_{200}^{\rm EdS}$ as used in the
empirical mass-luminosity relation of Reiprich \& B\"ohringer.

With the function $E(z)$ defined in Sect.\,\ref{MODEL}, a possible
redshift-dependence might be introduced by $\rho_c(z)\sim E^2(z)$ and
$\Delta_c(z)=18\pi^2+82\left[\Omega_m(z)-1\right]-39
\left[\Omega_m(z)-1\right]^2$ with
$\Omega_m(z)=\Omega_m(1+z)^3/E^2(z)$ for spatially flat geometries
(Bryan \& Norman 1998). For a given temperature we thus obtain
\begin{equation}\label{V9}
M_{\rm VIR}(z)\,=\,M_{\rm
VIR}(0)\,\,\frac{\Pi\left(\Delta_c(z),\Delta_c(0)\right)}{E(z)}\,,
\end{equation}
so that the relation between a virialized mass at redshift $z$ and the
mass as defined in the empirical mass/X-ray luminosity relation
becomes
\begin{equation}\label{V10}
M_{\rm VIR}(z)\,=\,M_{200}^{\rm EdS}(0)\,\cdot\,\Pi\left(\Delta_c(z),\Delta_c^{\rm
EdS}(0)\right)\,\cdot\,\frac{1}{E(z)}\,.
\end{equation}
Finally, we have to relate $M_{\rm VIR}(z)$ to the mass as defined in
the universal mass function of Jenkins et al. (2001). Here, the masses
are defined by the radius of the spherical overdensity of $18\pi^2$
with respect to the $\Omega_m$-dependent background density. Denoting
this mass by $M_{\rm SO(180)}(z)$ we have
\begin{equation}\label{V11}
M_{\rm SO(180)}(z)\,=\,M_{\rm VIR}(z)\,\,\Omega_m(z)\,\,\tilde{r}^3\,,
\end{equation}
where $\tilde{r}=r_{\rm SO(180)}/r_{\rm VIR}$ is obtained iteratively
from the NFW mass density profile using
\begin{equation}\label{V12}
\ln(1+c\tilde{r})-1+\frac{1}{1+c\tilde{r}}=\Omega_m(z)
\left[\ln(1+c)-\frac{c}{1+c}\right]\,.
\end{equation}
In the main text a non-evolving cluster population is assumed so that
the relevant equations are evaluated either at $z=0.0$ or
alternatively at $z=0.05$.

\end{document}